\documentstyle[preprint,epsf,aps]{revtex}
\begin{document}
\preprint{Draft : \today \ \ \ \ \ {\em [To be submitted to
Phys. Rev. C]}}

\title{Effects of the magnetic moment interaction between nucleons 
on observables in the 3N continuum}

\author{H. Wita{\l}a, J. Golak, R. Skibi\'nski}

\address{Insitute of Physics, Jagiellonian University,
Reymonta 4, PL-30059 Krak\'ow, Poland}

\author{C.R. Howell, W. Tornow}

\address{Physics Department, Duke University, Durham,
North Carolina, 27708
and Triangle Universities Nuclear Laboratory, Durham, North Carolina 27708}

\date{Received ---------2003}
\maketitle

\begin{abstract}
The influence of the magnetic moment interaction of nucleons on 
nucleon-deuteron elastic
scattering and breakup cross sections and on elastic scattering
polarization observables has been studied.
 Among the numerous elastic scattering observables only
the vector analyzing powers were found to show a significant effect, and of
opposite sign for the proton-deuteron  and neutron-deuteron systems.
This finding results in an even larger
discrepancy than the one previously established between neutron-deuteron  
data and theoretical calculations. For the
breakup reaction the largest effect was found
 for the final-state-interaction cross sections.
The consequences of this observation on previous determinations of 
the $^1S_0$ scattering lengths
from  breakup data are discussed.
\end{abstract}
\newpage

\vspace{.12in}

\section{Introduction}

The study of three-nucleon (3N) bound states and
reactions in the 3N continuum has improved
significantly our knowledge of the nuclear Hamiltonian~\cite{glo96,nog00}.
The underbinding of the triton
and $^3$He nuclei by modern nucleon-nucleon (NN) interactions was the 
first evidence for the necessity of including three-nucleon forces 
(3NF)~\cite{nog00} in addition to the pairwise NN interactions.
Furthermore, results of  Green function Monte-Carlo calculations~\cite{pie01}
showed that the energy levels of light nuclei can be explained only when the 
pairwise NN interactions are supplemented by appropriate 3NF's.
Additional evidence for 3NF effects came from the study of the cross-section 
minimum~\cite{wit98} in elastic nucleon-deuteron ($Nd$) scattering 
and the deuteron vector analyzing powers~\cite{sak00,cad01,sek02}.

Despite the spectacular successes obtained in interpreting 3N 
data based on the concept of a 3N Hamiltonian with free NN interactions and
supplemented by 3NF's,  some dramatic
discrepancies remain between theory and data that require further 
investigation.
These discrepancies can be divided into two categories according to energy.
One was discovered  at incident nucleon lab energies in the 3N system above
100~MeV and is exemplified by the  
nucleon-vector~\cite{cad01,bie00,erm01}
and deuteron-tensor
analyzing powers~\cite{sak00,wit01} in $Nd$ elastic scattering.
Since the 3NF effects become more important with increasing
energy~\cite{wit01,kur02},
these discrepancies will play an important role in establishing the
proper spin-isospin structure of the 3NF.
The second category was found at lab energies below 40 MeV. In the following
we will focus on these low-energy discrepancies.

The most famous is the vector analyzing power in $Nd$ elastic scattering.
The theoretical
predictions based on modern NN interactions, and including
3NF models, underestimate  considerably the maximum of the nucleon analyzing 
power
$A_y(\theta)$ in $\vec{p}$-$d$ and $\vec{n}$-$d$ scattering, as well as the
maximum of the deuteron vector analyzing power $iT_{11}(\theta)$ in 
$\vec{d}$-$p$ scattering~\cite{wit01,wit94,kie96}. 
At low energies (up to $\approx 30$~MeV)
 these two observables are very sensitive
 to changes in the $^3P_j$ NN and/or in $^4P_J$ $Nd$
phase shifts~\cite{tor98,tor02}. Even very small
changes of these phase shifts result in significant variations of 
$A_y(\theta)$ and
$iT_{11}(\theta)$.

Furthermore, low-energy neutron-deuteron ($nd$) breakup cross sections
show clear discrepancies between theory and
data~\cite{glo96,set96,str89,ste89,sst10p5,tunl99a,tunl99,krug??,siep02}
for some special kinematical arrangements of the outgoing three nucleons.
The most spectacular ones are the symmetrical
space-star  (SST) and the quasi-free scattering (QFS) configurations.

In the SST configuration the three nucleons emerge
in the c.m. system in a
plane perpendicular to the incoming beam with momenta of equal magnitudes
and directed such that the angle  between adjacent particle momentum vectors
is $120^{\circ}$. The low-energy $nd$ SST cross sections are clearly
underestimated by theoretical
predictions~\cite{set96,str89,ste89,sst10p5,tunl99a,tunl99}.
The calculated cross sections are insensitive to the NN potential
used in the calculations~\cite{glo96,kur02}.
They also do not change when any one of the present-day 3NF models is 
included~\cite{glo96,kur02}.

The QFS refers to the situation where one of the nucleons is at rest in the
laboratory system. In  $nd$ breakup,  $np$ or   $nn$ can form a
 quasi-free scattered pair.
Both cases have been measured~\cite{krug??,siep02}.
The picture resembles that for SST: the theoretical
 QFS cross sections are practically
independent of the NN potential used in calculations,
and they do not change when 3NF's are included~\cite{glo96,kur02}.
The calculated QFS scattering
cross section follows nicely the data when $np$ is the
 quasi-free interacting pair~\cite{siep02}.
However, when the $nn$ pair is quasi-free scattered instead of $np$,
 the theory clearly underestimates the experimental cross
sections~\cite{krug??,siep02}, similarly to the $nd$ SST case.

A problem of a different kind arises in the $nd$ breakup 
final-state-interaction 
(FSI) configuration where the two outgoing nucleons
have equal momenta.
The cross section for this geometry is characterized by a pronounced
peak when the relative energy of the final-state interacting
pair reaches zero (exact FSI condition). Due to the large sensitivity of this
enhancement to the $^1S_0$ NN scattering lengths, the FSI geometry has 
recently been studied in $nd$ breakup with the aim of determining the $nn$ 
$^1S_0$ scattering length~\cite{trot99,huh00}.
The $np$ FSI cross-section measurements performed simultaneously in 
\cite{trot99} and in two consecutive experiments in \cite{huh00,deng02} 
seem to indicate that indeed, this configuration
is a reliable tool for determining the NN $^1S_0$ scattering length: the 
values obtained for $a_{np}$ agreed with the result known from free $np$
scattering~\cite{anp}.  However, the values obtained for $a_{nn}$ 
in \cite{trot99,huh00} are in striking disagreement with each other, with the
result of \cite{trot99} in excellent agreement with the accepted value for
$a_{nn}$.

All previously published 3N continuum Faddeev calculations for the $nd$
system were restricted to pure strong nuclear forces, while for the $pd$ system 
the Coulomb interaction between the two protons had been included
in addition to the nuclear force~\cite{kie96}. However, the rigorous 
inclusion of the Coulomb interaction is currently limited to elastic
scattering.  More subtle electromagnetic contributions, such as the
magnetic moment interaction (MMI) between the nucleons have been
neglected in exact Faddeev calculations for the 3N continuum.
The approximate calculation by Stoks~\cite{sto98} at $E_N=3$~MeV, based
on a quasi two-body approach represents only the leading term of a genuine
3N calculation. Since this calculation predicted only a tiny effect on 
$A_y(\theta)$ in the region of the maximum it was generally concluded that
the exact treatment of the MMI was not worth the effort.  However, the
speculative interpretation \cite{nei02} of new data for $n$-$d$ scattering 
at very low energies and the associated comparison to $p$-$d$ scattering 
suggested that 
the MMI is indeed an important ingredient in the 3N continuum.

It is the aim of the present paper to go beyond the approximate
calculation referred to above and to study extensively the effects of the MMI 
on $Nd$ elastic scattering and breakup observables using the Faddeev approach.
Even though it is very unlikely that such subtle effects will have any 
significant influence on the QFS and SST cross sections, we found it 
worthwhile to calculate the magnitude of the MMI effects for these 
configurations.  In section~II we present the basic
theoretical  ingredients of our 3N calculations together with a short
description of the nucleon magnetic moment interactions.
The results for elastic scattering and breakup observables are 
presented and discussed in sections~III and IV, respectively.
In the breakup section we also focus on the consequences of the MMI 
effects on the extraction of the $^1S_0$ scattering lengths  from $nd$ breakup
data. Specifically, we present corrections induced by the MMI
on the values for $a_{np}$ and $a_{nn}$ deduced from very recent 
measurements. We summarize and conclude in section~V.

\section{Theoretical formalism}

The transition amplitudes for $Nd$ elastic scattering,
$<\Phi' \vert U \vert \Phi >$, and breakup, $<{\Phi}_0 \vert U_0 \vert \Phi >$,
can be expressed in terms of the vector $T\vert \Phi >$, which fulfills the 
3N Faddeev equation~\cite{glo96},

\begin{eqnarray}
\label{e1}
T \vert \Phi > &=& t P \vert \Phi > + t P G_0 \vert \Phi >,
\end{eqnarray}

as

\begin{eqnarray}
\label{e2}
< \Phi' \vert U \vert \Phi > &=& < \Phi'  \vert PG_0^{-1} + PT \vert \Phi >
 \cr
< {\Phi}_0 \vert U_0 \vert \Phi >
&=& < {\Phi}_0  \vert (1 + P) T \vert \Phi >~.
\end{eqnarray}

The incoming state $\vert \Phi > = \vert \overrightarrow q_0, \phi_d >$
is composed of the deuteron wave function $\phi_d$ and the momentum eigenstate
$\vert \overrightarrow q_0>$ of the relative nucleon-deuteron motion.
For elastic scattering the outgoing relative momentum changes its direction
leading to a state $\vert \Phi' >$, while for the breakup reaction
the final state $\vert \Phi_0 >$ is a momentum eigenstate that describes the
final motion of three outgoing  nucleons. The permutation operator P
takes into account the identity of nucleons and
it is the sum of a cyclic and anticyclic permutation of three particles.
The exchange term $PG_0^{-1}$ ($G_0$ denotes the free 3N propagator)
in the elastic
scattering amplitude results from the  interchange of the incoming nucleon
with those in the deuteron. All final-state interactions between the outgoing
nucleons, driven by a two-nucleon off-shell transition-matrix, that is denoted
as $t$ in Eqs.~\ref{e1} and \ref{e2}, are comprised in the
T operator, as can be easily seen by iterating  Eq.(\ref{e1}).

This formulation for nucleon-deuteron scattering assumes pairwise
interactions between  nucleons via  a short-range force V, which generates
through the Lippmann-Schwinger equation the transition-matrix $t$.
Therefore, it excludes the treatment of the long-range Coulomb force, but 
allows for the inclusion of any electromagnetic contribution of short-range 
character, such as, e.g., the magnetic moment interactions (MMI) between
nucleons.
In the case of our ``$pd$ calculations,'' the value of the magnetic moment 
of the 
two neutrons in our $nd$ breakup calculations is replaced by the value of the 
proton magnetic moment, i.e.,
the long-range Coulomb force is not included, neither is any interference 
between the Coulomb force and the MMI.

In our approach
we solve  Eq.(\ref{e1}) in momentum space and partial wave basis using
the magnitudes of the
standard Jacobi momenta $p=\vert \overrightarrow p \vert$ and
$q=\vert \overrightarrow q \vert$ to
describe the relative motion of the three nucleons, supplemented
by  angular momenta, spin, and isospin quantum numbers. 
Due to the short-range assumption, the result is a
finite set of coupled integral equations in two
continuous variables, $p$ and $q$.  The equations are solved for the 
amplitudes $< p q \alpha \vert T \vert \Phi >$ 
for each total angular momentum J and parity of the 3N system
by generating the Neumann series of
 Eq.(\ref{e1})  and summing it up by the Pade method.
The $\vert \alpha >$ is a set of angular, spin and isospin
quantum numbers
$\vert \alpha > \equiv \vert (ls)j (\lambda 1/2)I (jI)J (t1/2)T>$ ,
which describes the coupling of
the two-nucleon subsystem and the third nucleon to the
total angular momentum J and total isospin T of the 3N system.
  For details of the
theoretical formalism and numerical performance we refer to~\cite{glo96}.

To study effects of the MMI's of the nucleons we included in addition to the 
strong AV18~\cite{av18}  or CD~Bonn~\cite{cdb}
potentials the interactions of the magnetic moments in the $pp$, $np$, and
$nn$ subsystems.  The form and parametrization of the MMI's 
are given in Eqs.(8), (15), and (16) of ref.~\cite{av18}
(in the case of  the $np$ system the term
${\overrightarrow L}  {\overrightarrow A}$
in Eq.(15) of ref.~\cite{av18} was neglected).

The difference of the $pp$ ($nn$) and $np$ interactions in isospin t=1 
states induces transitions between  3N states with total isospin T=1/2 
and T=3/2. The strength of these transitions is
determined through the known charge-independence breaking of the NN
interactions~\cite{wit91}.  This charge-independence breaking 
can be treated approximately by a simple ``2/3-1/3 rule'',
for which  the effective t=1 transition matrix is
given by $t=2/3t_{pp(nn)} + 1/3t_{np}$, and
 T=3/2 3N states are neglected. This procedure
 is sufficient for most of the 3N scattering
observables~\cite{wit91}.
Since it is not evident that such an approximate approach is sufficient
when the MMI's are included, we performed also calculations where
 for each partial wave state
$\vert \alpha >$ with isospin t=1 both values of the total 3N isospin
T=1/2 and T=3/2 were taken into account. In all calculations we considered
all basis states $\vert \alpha >$ with two-nucleon subsystem angular
momenta up to $j \le j_{max}=3$.
In order to obtain full convergence of the numerical results for elastic 
scattering and breakup observables, Eq.(\ref{e1}) was
solved for total 3N angular momenta up to
$J=25/2$. Under this condition the maximal number of coupled integral
equations, which is equal to the number of possible $\vert \alpha >$'s,
amounts to  62 for the approximate
approach and increases to 89 when both values of the total isospin are taken
into account.

\section{Elastic scattering results}

\begin{figure}
\leftline{\mbox{\epsfysize=180mm \epsffile{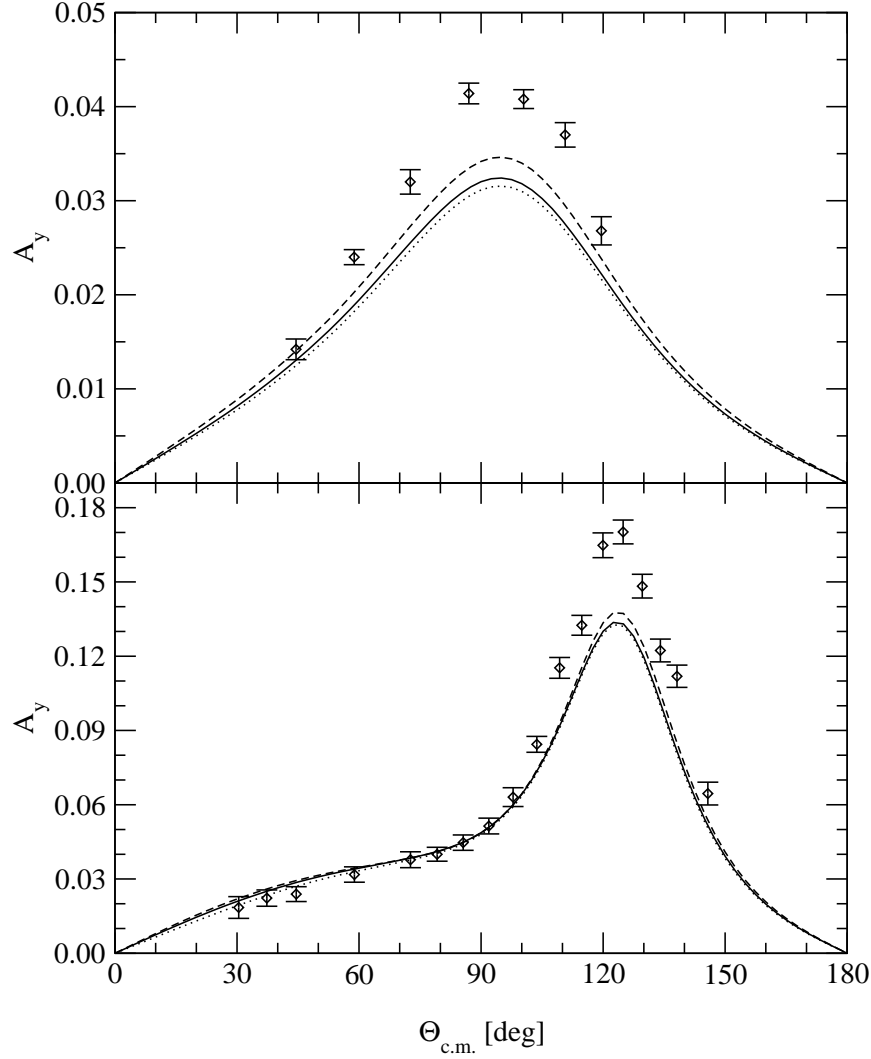}}}
\caption[]{
Effects of MMI's on $Nd$ elastic scattering vector analyzing power
$A_y(\theta)$ at $E_n^{lab}=1.9$~MeV (upper part) and $E_n^{lab}=10.0$~MeV 
(lower part).
The solid curve is the CD~Bonn potential prediction.
Inclusion of MMI's in the $pd$ system leads to the
dashed curve and for the $nd$ system to the dotted curve.
The $nd$ experimental data are from
\cite{nei02}
 ($1.9$~MeV) and
\cite{ay10}
($10.0$~MeV).
}
\label{fig:1a}
\end{figure}

\begin{figure}
\leftline{\mbox{\epsfysize=180mm \epsffile{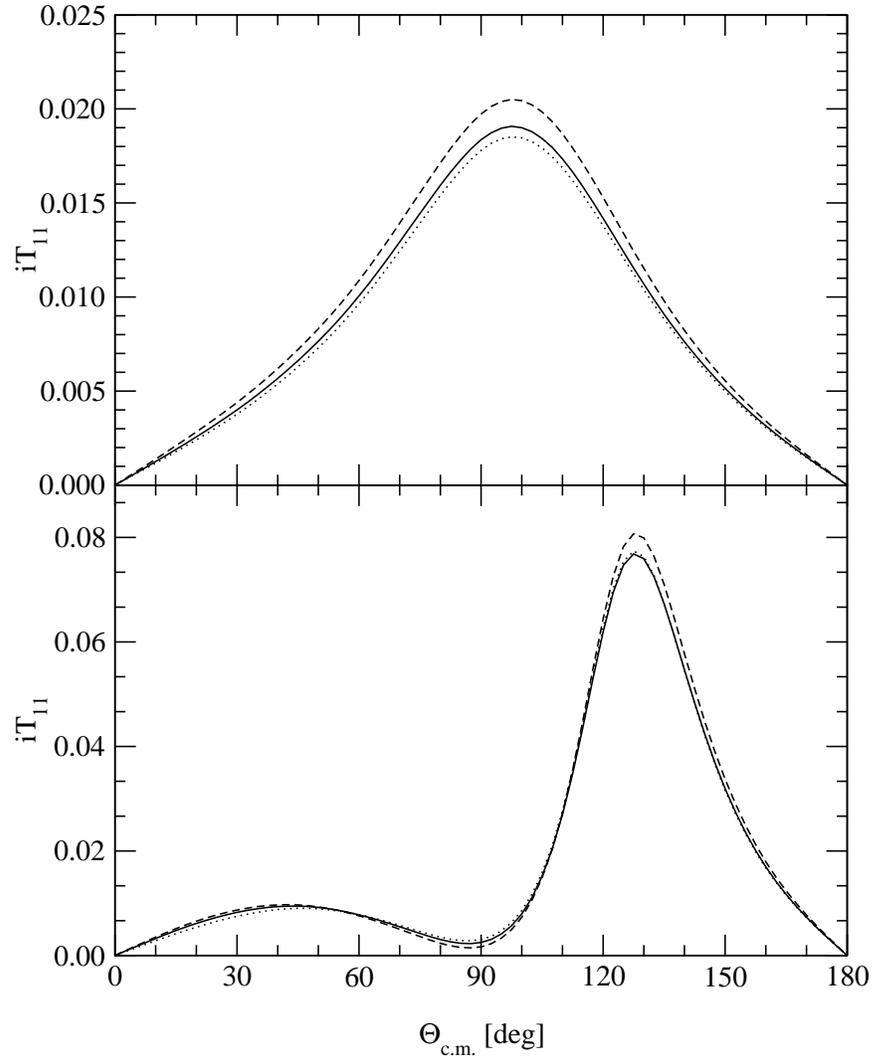}}}
\caption{Effects of MMI's on $Nd$ elastic scattering tensor analyzing power
$iT_{11}(\theta)$ at $E_n^{lab}=1.9$~MeV (upper part) and $E_n^{lab}=10.0$~MeV
(lower part).
The solid curve is the CD~Bonn potential prediction. Inclusion of MMI's in 
the  $pd$ system leads to
the dashed curve and for the $nd$ system to the dotted curve.
}
\label{fig:1b}
\end{figure}

\begin{figure}
\leftline{\mbox{\epsfysize=180mm \epsffile{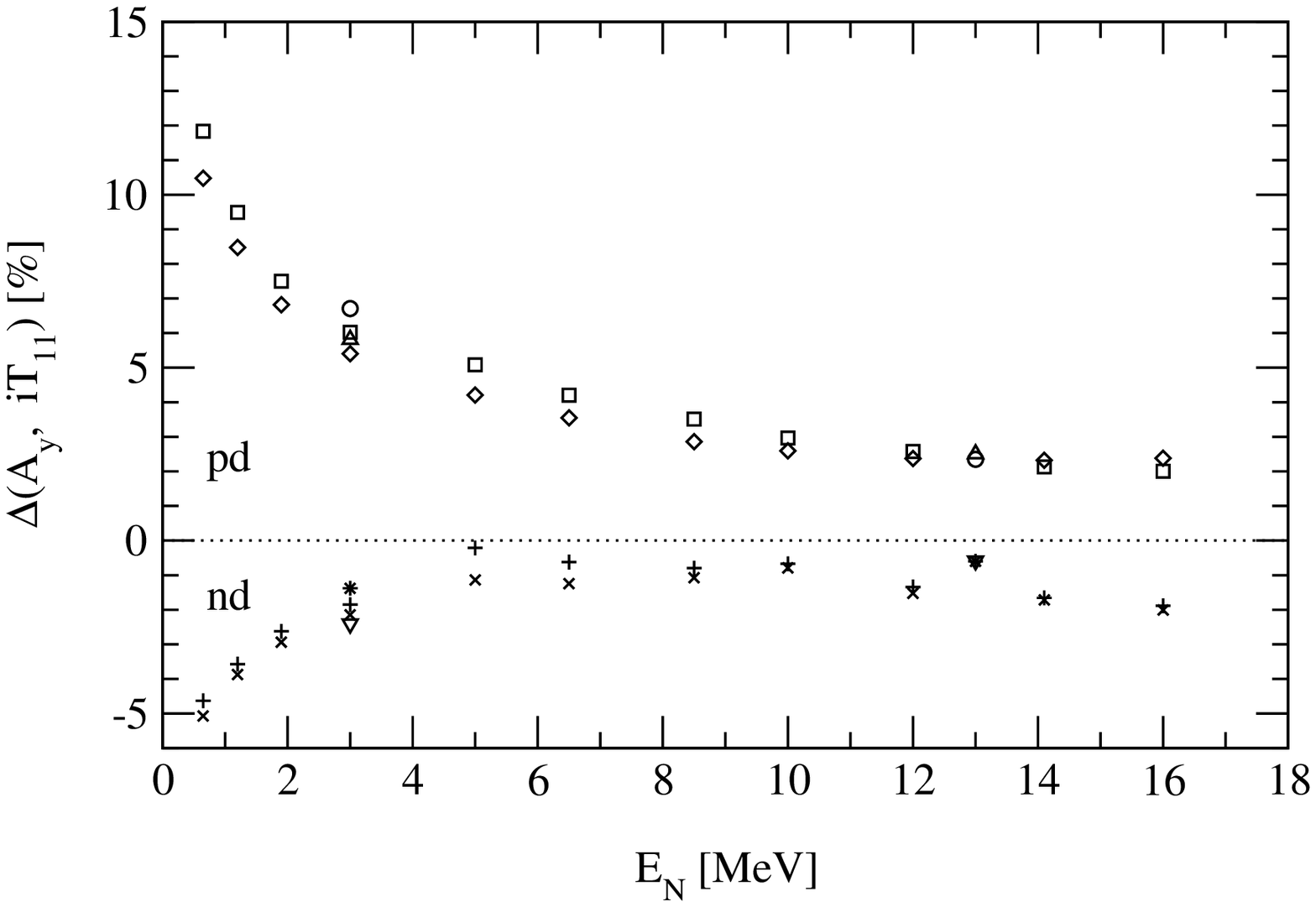}}}
\caption{The energy dependence of the MMI's effect $\Delta=
{ {[A_y^{(NN)}(iT_{11}) -A_y^{(NN+MMI's)}(iT_{11})] }\over
{A_y^{(NN)}(iT_{11})} } * 100\%$ on vector analyzing
powers for elastic $Nd$ scattering at the angle where these observables reach a
maximum value.
The diamonds and pluses represent $\Delta$ predicted by the CD~Bonn
potential for the $pd$ and $nd$ $A_y$, respectively. The squares and crosses 
represent $\Delta$ for the $pd$ and $nd$
$iT_{11}$, respectively. Also, at 3 MeV and 13 MeV the AV18 results are 
shown (triangle up (star) for the $pd$ ($nd$) $A_y$, and
circle (triangle down) for the $pd$ ($nd$) $iT_{11}$,
respectively).
}
\label{fig:2}
\end{figure}

The nucleon scattered off the deuteron can be either a neutron or a proton. 
Since they have magnetic moments of different sign and magnitude, we 
performed separate calculations for the $pd$ and $nd$ systems. In both cases 
the
$pp$ ($nn$)  and $np$ nuclear interactions of the AV18 and CD~Bonn potentials 
were used, supplemented by the appropriate MMI's: $pp$ and $np$ for the $pd$ 
system, and $nn$ and $np$ for the $nd$ system.
Comparisons of the theoretical predictions
 for elastic scattering observables were made between calculations 
obtained with and without the
MMI's  included. Among the unpolarized
cross section, analyzing powers, spin correlation coefficients,
 and polarization transfer coefficients,
only the vector analyzing powers [$A_y(\theta)$ and $iT_{11}(\theta)$]
show a significant influence of the MMI's (see Fig.~\ref{fig:1a}, 
~\ref{fig:1b}, and~\ref{fig:2}).
As expected from the different signs of the proton and neutron magnetic 
moments, the effects of the MMI's have opposite signs for the $pd$ and $nd$ 
systems.

For the $pd$ system the MMI's raise the maximum value of $A_y(\theta)$ and 
$iT_{11}(\theta)$ by $\approx 4\%$ at $E_n^{lab}=3$~MeV,
thus bringing it closer to the experimental $pd$ data, while for
the $nd$ system they  reduce the maximum of
$A_y(\theta)$ and $iT_{11}(\theta)$ by $\approx 3\%$, therefore enlarging the
 discrepancy between theory and data. The magnitude of the MMI
effects is energy dependent and decreases with increasing nucleon
energy (see Fig.~\ref{fig:2}). The contribution of the MMI's is
thus most significant at low energies.
The relative magnitude of the effect is comparable for
 $A_y(\theta)$  and $iT_{11}(\theta)$ and
is roughly independent of the strong NN interaction
used in the calculations.

We found that the evaluation of the effects induced
by the MMI's does not require partial wave components with 
 total isospin T=3/2. Restricting the calculations to the
approximate ``2/3-1/3 rule''
 leads to a fairly good estimate of the MMI effects.

Our results once again exemplify the spectacular sensitivity of the 
low-energy vector analyzing powers to the $^3$P-wave interactions. 
In spite of the relative
 smallness of the MMI contributions  to the potential
energy of the three nucleons, their effect is amplified by this $^3$P-wave
sensitivity, and they must be taken into account in any
final solution to the $A_y(\theta)$ puzzle.

Recent experimental data for the low-energy $nd$ $A_y(\theta)$ 
($E_n=1.2$ and $1.9$~MeV) revealed a sizable difference
with respect to $pd$ data. The difference increases with
decreasing center-of-mass energy~\cite{nei02}. This energy dependence
was used in \cite{nei02} to speculate that the difference between the $
nd$ and $pd$ $A_y(\theta)$ data at low energies is due to the MMI's
of the three nucleons in the 3N continuum.  The present work clearly supports
this conjecture.

\section{Breakup results}

The final deuteron breakup  state requires 5 independent kinematical
parameters to define it unambiguously.
They can be taken as, e.g., laboratory energies $E_1$ and $E_2$ of two 
outgoing nucleons together with their angles to define
the directions of their  momenta. In order to locate regions in this
 5-dimensional breakup phase space with large changes of the cross section
due to the MMI's of the nucleons, we applied the
projection procedure described in details
in ref.~\cite{kur02}. The magnitude of the MMI effects
on the exclusive breakup cross section
${{d^5{\sigma}}\over{d{\Omega}_1d{\Omega}_2dS} }$ along the kinematically 
allowed S-curve was defined as
\begin{eqnarray}
\label{e3}
\Delta &=& \vert {{  {{d^5{\sigma}^{(noMMI)} }
\over{d{\Omega}_1d{\Omega}_2dS} } -
{{d^5{\sigma}^{(MMI)} }\over{d{\Omega}_1d{\Omega}_2dS} }   }
\over{ {{d^5{\sigma}^{(MMI)} }
\over{d{\Omega}_1d{\Omega}_2dS} }  }}  \vert * 100\%.
\end{eqnarray}
We searched the entire breakup phase space for the distribution of
$\Delta$ values. For this purpose the phase space was projected onto 
three sub-planes:
 $\theta_1-\theta_2$, $\phi_{12}-\theta_2$, and $E_1-E_2$. Here, 
$\theta_1$ and $\theta_2$ are the polar angles
which together with the azimuthal angles $\phi_1$ and $\phi_2$
($\phi_{12}=\phi_1 -\phi_2$) define the directions of the nucleon momenta.
 The resulting projections of $\Delta$  are shown in Fig.~\ref{fig:3}
at the three incoming proton energies of $E_p^{lab}=5, 13$ and $65$~MeV for 
the $^2$H$(p,pp)n$ breakup reaction.
The largest changes found for the breakup cross sections
reach $\approx 10\%$, and even at 65~MeV there are configurations
with non-negligible MMI effects. The largest effects are located mostly
in  regions of the phase space that are characterized by
FSI geometries. In most other parts of the phase space the effects
of the MMI's are rather small. In particular, the SST and QFS  cross
sections are only slightly influenced. According to
our calculations, for the QFS configuration the MMI effect depends on the
laboratory angles of the quasi-free interacting nucleons and it is
different for the $pp$, $nn$, or $np$ pairs. However, the
effects are quite small. At $E_p^{lab}=13$~MeV they are never  larger than
$2\%$. For the SST configuration the effects are smaller than $1\%$. 
Changing the c.m. angle between the space-star plane and the beam axis to 
values different from $90^{\circ}$ results in only small changes of the 
cross section of less than 2\% when the  MMI's are included.

\begin{figure}
\leftline{\mbox{\epsfysize=180mm \epsffile{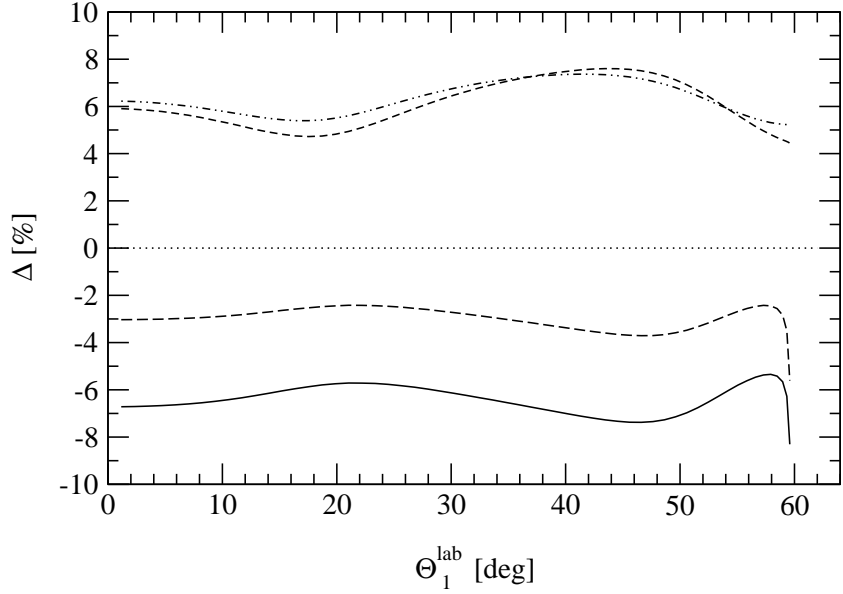}}}
\caption{Effects of MMI's $\Delta$ obtained with the CD~Bonn potential, on the
  $E_n^{lab}=13$~MeV  
FSI cross sections as a function of the production angle $\theta_1^{lab}$.
The solid and  long-dashed curves correspond to the $pp$ FSI in the
reaction $^2$H$(p,pp)n$ and the $nn$ FSI in the reaction $^2$H$(n,nn)p$, 
respectively. The effects on the  $np$ FSI in these
breakup reactions are given by the short-dashed and dashed-double-dotted 
curves.}
\label{fig:6}
\end{figure}

\begin{figure}
\leftline{\mbox{\epsfysize=180mm \epsffile{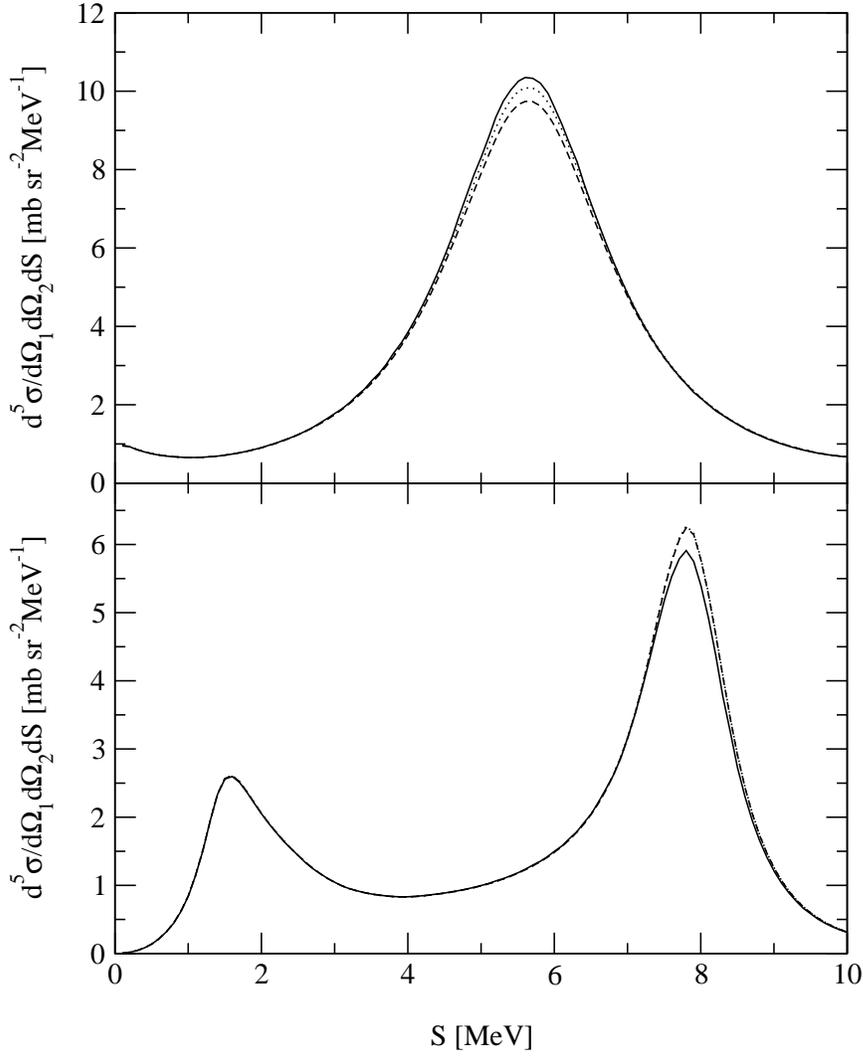}}}
\caption[]{
MMI effects on $E_n^{lab}=13$~MeV $nd$ breakup $nn$ FSI [$^2$H$(n,nn)p$,
$\theta_1=\theta_2=28^{\circ}, \phi_{12}=0^{\circ}$] (upper part) and $np$ FSI
[$^2$H$(n,np)n$, 
$\theta_1=28^{\circ}, \theta_2=83.5^{\circ}, \phi_{12}=180^{\circ}$]
(lower part) configurations of ref.~\cite{trot99}.
The CD~Bonn potential prediction is
 given by the solid curve,  and the dashed (dotted) curves refer to the
calculations with the $pp$-$np$ ($nn$-$np$) MMI's included.
Note the overlapping results for the $pp$-$np$ and $nn$-$np$ MMI's in the 
$np$ FSI peak.
}
\label{fig:7}
\end{figure}

\begin{figure}
\leftline{\mbox{\epsfysize=180mm \epsffile{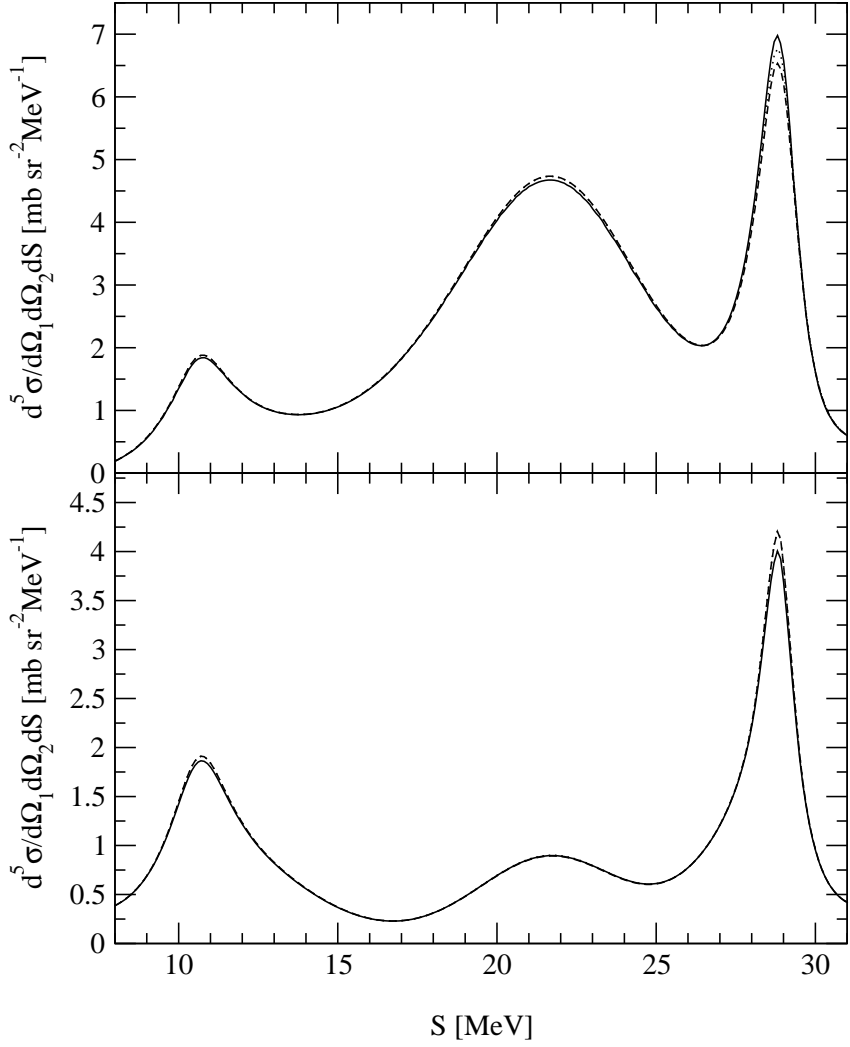}}}
\caption[]{
MMI effects in $E_n^{lab}=25.3$~MeV $nd$ breakup $nn$ FSI [$^2$H$(n,np)n$,
 $\theta_1=55.5^{\circ}$, $\theta_2=41.15^{\circ}$, $\phi_{12}=180^{\circ}$]
 (upper part)
 and $np$ FSI
[$^2$H$(n,nn)p$, $\theta_1=55.5^{\circ}, \theta_2=41.15^{\circ},
 \phi_{12}=180^{\circ}$] (lower part)
 configurations of ref.~\cite{huh00}.
 Descriptions of curves are the same as in Fig.~\ref{fig:7}.
 Note the overlapping  results for the $pp$-$np$ and $nn$-$np$ MMI's in the 
$np$ FSI peak.
}
\label{fig:8}
\end{figure}

As stated above, the largest changes of the cross sections of up to 
$\approx 10\%$ occur for FSI configurations. Their relative magnitude
depends on the incoming beam energy and on the production angle of the
final-state interacting pair (the lab. angle between the
 momentum of the FSI pair and the beam axis).
The sign of the effect depends on the type of the final-state interacting 
pair (see Figs.~\ref{fig:6}-\ref{fig:8}).
For the $np$ FSI peak the cross section is increased and for the $pp$ ($nn$) 
FSI peak it is decreased, according to the increased or decreased attraction
caused by the MMI's in the corresponding NN subsystem. The relative
changes of the FSI  cross sections are
comparable for the $np$ and $pp$ FSI peaks, and are a factor of
$\approx 2$ smaller for the $nn$ FSI peak (see Fig.~\ref{fig:6}).
This factor approximately corresponds to the ratio of the squares of 
the neutron and proton magnetic moments, $\mu_p^2 / \mu_n^2 = 2.13$.

This behavior of the FSI cross sections is of interest in view
of recently reported values for the $^1S_0$ $nn$ and $np$ scattering lengths
extracted from $nd$ breakup measurements~\cite{trot99,huh00}.
As stated earlier, for the $np$
system both measurements resulted in comparable values for $a_{np}$, which 
agree with the value obtained from free $np$ scattering. However, the reported
results for the $a_{nn}$ scattering
length are strikingly different. In Fig.~\ref{fig:7}
 we show the $nn$ and $np$ cross sections for the FSI configurations of 
ref.~\cite{trot99} and in Fig.~\ref{fig:8} for the corresponding configurations
of ref.~\cite{huh00}, together with the effects induced by the MMI's on the
FSI peaks.
It is interesting to note that the theoretical cross sections for the $np$ FSI
obtained with MMI's included do not change significantly
if the $np$ pair is accompanied by either a $nn$ or $pp$ pair. This effect is 
slightly dependent on the production angle of the FSI pair 
(see Fig.~\ref{fig:6}).
This observation, together with our findings about the dependence of the 
$nn$ FSI peak on the magnitude of the MMI's, indicates
a  relatively simple mechanism by which the MMI's affect FSI geometries, 
resulting in a net effect that is dominated by the
magnetic moments of the FSI pair.

In view of the non-negligible effects of the MMI's on FSI cross sections
found in the present work one would conclude that the values reported 
in refs.~\cite{trot99}, ~\cite{huh00}, and ~\cite{deng02}
for the scattering lengths must be corrected for such effects.
Based on the sensitivity of the theoretical point-geometry FSI cross sections
to specific values of $a_{np}$ and $a_{nn}$ for the geometries of 
the experiments described in ~\cite{trot99},
~\cite{huh00}, and ~\cite{deng02},
the MMI associated corrections for the $a_{np}$ and $a_{nn}$ scattering
lengths are  shown in column 3 of Tables~\ref{tab:1} and ~\ref{tab:2},
respectively.
While the correction to the $nn$ scattering length is rather small, 
the correction to the $np$ scattering length moves the $a_{np}$ values 
obtained from the $nd$ breakup reaction away from the
free $np$ scattering result by $\approx 1$~fm. The corrections given in
Tables ~\ref{tab:1} and ~\ref{tab:2} are for point geometry, i.e., they do not
include the finite geometry of the experimental setup and the associated
energy smearing.

Such sizeable corrections, especially for $a_{np}$, if true, would cast 
doubt on the accuracy of previous results obtained from the $nd$
breakup reaction. Evidence that this is not the case is shown in 
Table~\ref{tab:3}.  This table contains the effects of including the
MMI's with the AV18 and the CD Bonn NN potentials in 2N calculations of the 
$^1$S$_0$ scattering
length.  The inclusion of the MMI's increases the magnitude of $a_{np}$, while
for the $pp$ and $nn$ systems, it decreases the magnitude of the scattering 
lengths. The magnitudes of changes
for $a_{np}$ and $a_{nn}$ are nearly equal and opposite in sign to
the corrections found for the 3N system (see column 3 in
Tables ~\ref{tab:1} and ~\ref{tab:2} and column 5 in Table ~\ref{tab:3}).
Therefore, the effective corrections, given by their sum, are nearly 
negligible. The final corrected values for $a_{np}$ and $a_{nn}$ are given
in column 4 of Tables ~\ref{tab:1} and ~\ref{tab:2}, respectively. 
Clearly, the  MMI's  do not explain the 
different values for $a_{nn}$ obtained in the measurements of ~\cite{trot99}
and ~\cite{huh00}.  As a side remark one should mention a shortfall of the 
$np$ $^1$S$_0$ scattering length used in the CD Bonn potential.  It was fitted
to the experimental value of $a_{np}$ obtained from free $n$-$p$ scattering
without taking the MMI into account.

\section{Summary and conclusions}

We performed an extensive study of effects induced by the magnetic moment 
interactions of nucleons in the 3N continuum.
For elastic $Nd$ scattering we found that only the vector
analyzing powers show significant changes when the MMI's are included.
For the $nd$ system the MMI's increase the discrepancy between calculations and
data for $A_y(\theta)$ in the region of the $A_y(\theta)$ maximum, while for 
the $pd$ system they reduce the discrepancy.
The effects for $iT_{11}$ are similar. The relative magnitude of the MMI
effects
decreases with increasing energy. The recent low-energy $nd$ $A_y(\theta)$ data
support the action of the MMI \cite{nei02}. Our results show that any final 
solution of the $A_y(\theta)$ puzzle must incorporate the MMI's of the
nucleons involved.

For the $Nd$ breakup cross sections the regions where the MMI's are important
are restricted to FSI
geometries, where changes of the cross sections of up to $\approx 10\%$
were found.
Such large effects originate from modifications of the $^1S_0$
scattering lengths caused by the MMI's, to which the FSI cross sections are
sensitive. Due to these modifications 
 when extracting scattering length from breakup measurement  
 the MMI's can be neglected in the underlying
theory, i.e., the resulting effective corrections are practically
negligible. 
Therefore, the very minor MMI corrections found in the present work
do not explain the different values for the $^1S_0$
$a_{nn}$ obtained in recent $nd$ breakup experiments.
From the theoretical point of view, the $^1$S$_0$ scattering lengths
$a_{nn}$ and $a_{np}$ can be reliably determined from experimental data, 
provided the MMI's are treated consistently on the NN {\it and} 3N level.
If the MMI's are included in only one of these levels, the associated results
for the scattering lengths must be corrected appropriately.

In most regions  of the phase space $Nd$ breakup 
cross sections are influenced only slightly by the MMI's. In particular,
this is the case for the QFS and SST configurations. Therefore, the MMI's 
are not  responsible for the large differences found between theoretical 
predictions and data  for the $nd$ breakup SST and $nn$~QFS cross sections.

\begin{acknowledgements}
This work was supported in part by the U.S. Department of Energy, Office of
High-Energy and Nuclear Physics, under grant No. DE-FG02-97ER41033, and
by the Polish Committee for Scientific Research
under grant No. 2P03B02818.
R.S thanks the Foundation for Polish Science for financial support.
The numerical calculations have been performed
on the Cray T90 of the Neumann Institute for Computing (NIC)
at the Forschungszentrum in J\"ulich,
Germany.
\end{acknowledgements}

\begin{table}
  \caption{The CD~Bonn corrections  to  $a_{np}$
 values extracted from $np$ FSI cross sections in $nd$ breakup
experiments (see text for explanation of columns).}

  \begin{tabular}{|c|r|r|r|} \hline
      {Configuration} & extracted  & MMI corrections & corrected \\
                      &  $a_{np}$ (fm) & (fm)  &  $a_{np}$ (fm) \\
\hline
      $\theta_1=28^\circ$, $\theta_2=83.5^\circ$,
$\phi_{12}=180^{\circ~[a]}$  & -23.7 $\pm$ 0.3~$^{[c]}$ & +0.947
& -23.7$\pm$ 0.3\\
      $\theta_1=35.5^\circ$, $\theta_2=69^\circ$,
$\phi_{12}=180^{\circ~[a]}$  & -23.2 $\pm$ 0.3~$^{[c]}$ & +0.985
& -23.2$\pm$ 0.3\\
      $\theta_1=43^\circ$, $\theta_2=55.7^\circ$,
$\phi_{12}=180^{\circ~[a]}$  & -23.6 $\pm$ 0.3~$^{[c]}$ & +1.006
& -23.7$\pm$ 0.3\\
      $\theta_1=55.5^\circ$, $\theta_2=41.15^\circ$,
$\phi_{12}=180^{\circ~[a]}$  & -23.9 $\pm$ 1.0~$^{[d]}$ & +0.955
& -23.9$\pm$ 1.0\\
      $\theta_1=\theta_2=32^\circ$,
$\phi_{12}=0^{\circ~[b]}$  & -24.3 $\pm$ 1.1~$^{[e]}$ & +0.908
& -24.3$\pm$ 1.1\\
\hline
  \end{tabular}
[a] $^2$H($n,nn)p$; [b] $^2$H($n,np)n$; [c] ref.~\cite{trot99} ($E_n^{lab}=13$~MeV);
[d] ref.~\cite{huh00} ($E_n^{lab}=25.3$~MeV);
[e] ref.~\cite{deng02} ($E_n^{lab}=25$~MeV).
  \label{tab:1}
\end{table}

\begin{table}
  \caption{The CD~Bonn corrections  to  $a_{nn}$
 values extracted from $nn$ FSI cross sections in $nd$ breakup
experiments (see text for explanation of columns).}

  \begin{tabular}{|c|r|r|r|} \hline
      {Configuration} & extracted  & MMI corrections & corrected \\
                      &  $a_{nn}$ (fm) & (fm)  &  $a_{nn}$ (fm) \\
\hline
      $\theta_1=\theta_2=20.5^\circ$,
$\phi_{12}=0^{\circ~[a]}$  & -18.9 $\pm$ 0.2~$^{[c]}$ & -0.244
& -18.8$\pm$ 0.2\\
      $\theta_1=\theta_2=28^\circ$,
$\phi_{12}=0^{\circ~[a]}$  & -18.8 $\pm$ 0.2~$^{[c]}$ & -0.254
& -18.7$\pm$ 0.2\\
      $\theta_1=\theta_2=35.5^\circ$,
$\phi_{12}=0^{\circ~[a]}$  & -17.7 $\pm$ 0.4~$^{[c]}$ & -0.279
& -17.6$\pm$ 0.4\\
      $\theta_1=\theta_2=43^\circ$,
$\phi_{12}=0^{\circ~[a]}$  & -18.8 $\pm$ 0.4~$^{[c]}$ & -0.306
& -18.7$\pm$ 0.4\\
      $\theta_1=55.5^\circ, \theta_2=41.15^\circ$,
$\phi_{12}=180^{\circ~[b]}$  & -16.1 $\pm$ 0.4~$^{[d]}$ & -0.318
& -16.1$\pm$ 0.4\\
      $\theta_1=55.5^\circ, \theta_2=41.15^\circ$,
$\phi_{12}=180^{\circ~[b]}$  & -16.2 $\pm$ 0.3~$^{[d]}$ & -0.303
& -16.1$\pm$ 0.3\\
\hline
  \end{tabular}
[a] $^2$H($n,nn)p$; [b] $^2$H($n,np)n$; [c] ref.~\cite{trot99} ($E_n^{lab}=13$~MeV);
[d] ref.~\cite{huh00} ($E_n^{lab}=25.3$~MeV);
[e] ref.~\cite{huh00} ($E_n^{lab}=16.6$~MeV);
.
\label{tab:2}
\end{table}

\begin{table}
  \caption{The experimental, AV18  and CD~Bonn potential $^1S_0$
scattering lengths, together with MMI corrections
(see text for explanation of columns).}

  \begin{tabular}{|c|r|r|r|r|} \hline
      system & Experiment  & NN potential & NN potential  &
MMI corrections \\
             &  (fm)         & (fm)       &  with MMI's (fm)  & (fm)     \\
\hline
       $np$ & -23.749 $\pm$ 0.008$^{[a]}$& AV18: -23.084 & -23.863$^{[d]}$
& -0.779    \\
          &                     & CD~Bonn: -23.740 &  -24.685   & -0.945  \\
       $pp$ &  -7.8063 $\pm$ 0.0026$^{[b]}$ & AV18: -17.164 & -16.581
 & +0.583  \\
          &                       & CD~Bonn: -17.466 &  -16.797  & +0.669  \\
       $nn$ &  -18.5 $\pm$ 0.4$^{[c]}$ & AV18: -18.818 & -18.487    & +0.331 \\
          &                  & CD~Bonn: -18.801 & -18.433   & +0.368  \\
\hline
  \end{tabular}
[a] ref.~\cite{anp}; [b] ref.~\cite{app}; [c] ref.~\cite{ann};
[d] in ~\cite{av18} $V_{C1}$ (Eq.~(12) in \cite{av18}) was included to get
experimental $a_{np}$ value.
\label{tab:3}
\end{table}


\begin{figure}
\caption{Projections of the relative MMI effects $\Delta$
on two-dimensional planes for $^2$H$(p,pp)n$ breakup at $E_p^{lab}=5, 13,$ 
and $65$~MeV (upper row, center row, and bottom row) obtained with the CD~Bonn potential. The color scale, adjusted separately for each energy, gives
the magnitude of the effect.
}
\label{fig:3}
\end{figure}

\newpage

\end{document}